\documentclass[conference]{IEEEtran}
\IEEEoverridecommandlockouts
\usepackage{cite}
\usepackage{amsmath,amssymb,amsfonts}
\usepackage{algorithmic}
\usepackage{graphicx}
\usepackage[american]{circuitikz}
\usepackage{textcomp}
\usepackage{xcolor}
\usepackage{makecell}
\usepackage{todonotes}
\usepackage{subcaption}
\usepackage{float}
\usepackage{etoolbox}
\newbool{showComments}
\booltrue{showComments}

\ifbool{showComments}{%

}{

}

\usepackage[all]{background}

\usepackage{stackengine}
\setstackEOL{\\}
\setstackgap{L}{\normalbaselineskip}
\SetBgContents{\color{gray}{\tiny \Longstack{PREPRINT - accepted at ACM/IEEE International Symposium on Low Power Electronics and Design 2023 (ISLPED 2023)}}}
\SetBgPosition{4.3cm,1cm}
\SetBgOpacity{1.0}
\SetBgAngle{0}
\SetBgScale{1.8}

\usepackage[a4paper, total={184mm,239mm}]{geometry}
\def\BibTeX{{\rm B\kern-.05em{\sc i\kern-.025em b}\kern-.08em
    T\kern-.1667em\lower.7ex\hbox{E}\kern-.125emX}}
\usepackage{comment}

\begin{document}
\bstctlcite{IEEEexample:BSTcontrol}

\tikzset{
  declare function={
    atan3(\a,\b)=ifthenelse(atan2(0,1)==90, atan2(\a,\b), atan2(\b,\a));},
  kinky cross radius/.initial=+.125cm,
  @kinky cross/.initial=+, kinky crosses/.is choice,
  kinky crosses/left/.style={@kinky cross=-},kinky crosses/right/.style={@kinky cross=+},
  kinky cross/.style args={(#1)--(#2)}{
    to path={
      let \p{@kc@}=($(\tikztotarget)-(\tikztostart)$),
          \n{@kc@}={atan3(\p{@kc@})+180} in
      -- ($(intersection of \tikztostart--{\tikztotarget} and #1--#2)!%
             \pgfkeysvalueof{/tikz/kinky cross radius}!(\tikztostart)$)
      arc [ radius     =\pgfkeysvalueof{/tikz/kinky cross radius},
            start angle=\n{@kc@},
            delta angle=\pgfkeysvalueof{/tikz/@kinky cross}180 ]
      -- (\tikztotarget)}}}

\ctikzset{tripoles/mos style/arrows}

\setlength{\belowcaptionskip}{-11pt} 



\title{IMBUE: \underline{I}n-\underline{M}emory \underline{B}oolean-to-C\underline{U}rrent Inference Architectur\underline{E} for Tsetlin Machines \vspace{-5mm}}
\author{\IEEEauthorblockN{Omar Ghazal\IEEEauthorrefmark{1}\IEEEauthorrefmark{5}, Simranjeet Singh\IEEEauthorrefmark{2}\IEEEauthorrefmark{6}, Tousif Rahman\IEEEauthorrefmark{1}, Shengqi Yu\IEEEauthorrefmark{1},
Yujin Zheng\IEEEauthorrefmark{1}, \\
Domenico Balsamo\IEEEauthorrefmark{1}, 
Sachin Patkar\IEEEauthorrefmark{2},
Farhad Merchant\IEEEauthorrefmark{1},
Fei Xia\IEEEauthorrefmark{1},
Alex Yakovlev\IEEEauthorrefmark{1},
Rishad Shafik\IEEEauthorrefmark{1}
\IEEEauthorblockA{\IEEEauthorrefmark{1}Newcastle University, UK,\IEEEauthorrefmark{2}Indian Institute of Technology Bombay, India,\\ \IEEEauthorrefmark{5}University of Mosul, Iraq,    \IEEEauthorrefmark{6}Forschungszentrum Jülich GmbH, Germany,}}
\{simranjeet, patkar\}@ee.iitb.ac.in,\{O.G.G.Awf2, s.rahman, y.zheng26, s.yu10, domenico.balsamo, \\ farhad.merchant, fei.xia, alex.yakovlev, rishad.shafik\}@newcastle.ac.uk \vspace{-7mm}}

\maketitle
\begin{abstract}
In-memory computing for Machine Learning (ML) applications remedies the von Neumann bottlenecks by organizing computation to exploit parallelism and locality. Non-volatile memory devices such as Resistive RAM (ReRAM) offer integrated switching and storage capabilities showing promising performance for ML applications. However, ReRAM devices have design challenges, such as non-linear digital-analog conversion and circuit overheads. This paper proposes an In-Memory Boolean-to-Current Inference Architecture (IMBUE) that uses ReRAM-transistor cells to eliminate the need for such conversions. IMBUE processes Boolean feature inputs expressed as digital voltages and generates parallel current paths based on resistive memory states. The proportional column current is then translated back to the Boolean domain for further digital processing. The IMBUE architecture is inspired by the Tsetlin Machine (TM), an emerging ML algorithm based on intrinsically Boolean logic. 
The IMBUE architecture demonstrates significant performance improvements over binarized convolutional neural networks and digital TM in-memory implementations, achieving up to a \emph{12.99}x and \emph{5.28}x increase, respectively.

\end{abstract}



\section{Introduction}
Machine Learning (ML) applications are intrinsically data-centric. This creates a compelling case for transitioning from traditional computing systems towards systems utilizing In-Memory Computing (IMC). IMC introduces the advantage of computations being performed within the memory itself~\cite{Sebastian2020}. This reduces latency and energy overheads from data movement, the so-called von Neumann bottleneck, and thereby facilitates the better acceleration of ML problems in resource-constrained micro-edge applications~\cite{Wang2022, Xia2019}. 


For effective IMC, typically, the memory compute unit is organized to leverage parallelism from the ML algorithm through crossbar-like structures. These crossbars capitalize on performance by exploiting the low power consumption and fast access times of their constituent memory devices~\cite{Liu2022,Wang2022,Xia2019}.

 Emerging analog memory devices such as Resistive RAM (ReRAM) present one promising approach to addressing drawbacks with the continued scaling down of CMOS-based memory devices. ReRAM offers non-volatile storage with high-density crossbars, reduced leakage currents, and memory size overheads~\cite{Yu2018, Sebastian2020}.
 However, one essential challenge with ReRAM devices is the nonlinearity in their operation within memory arrays. These variations can be manifested between devices and between cycles for the same device~\cite{Xia2019, Ni2017}.

\begin{figure}[!t]
    \centering
    \includegraphics[width=0.488\textwidth]{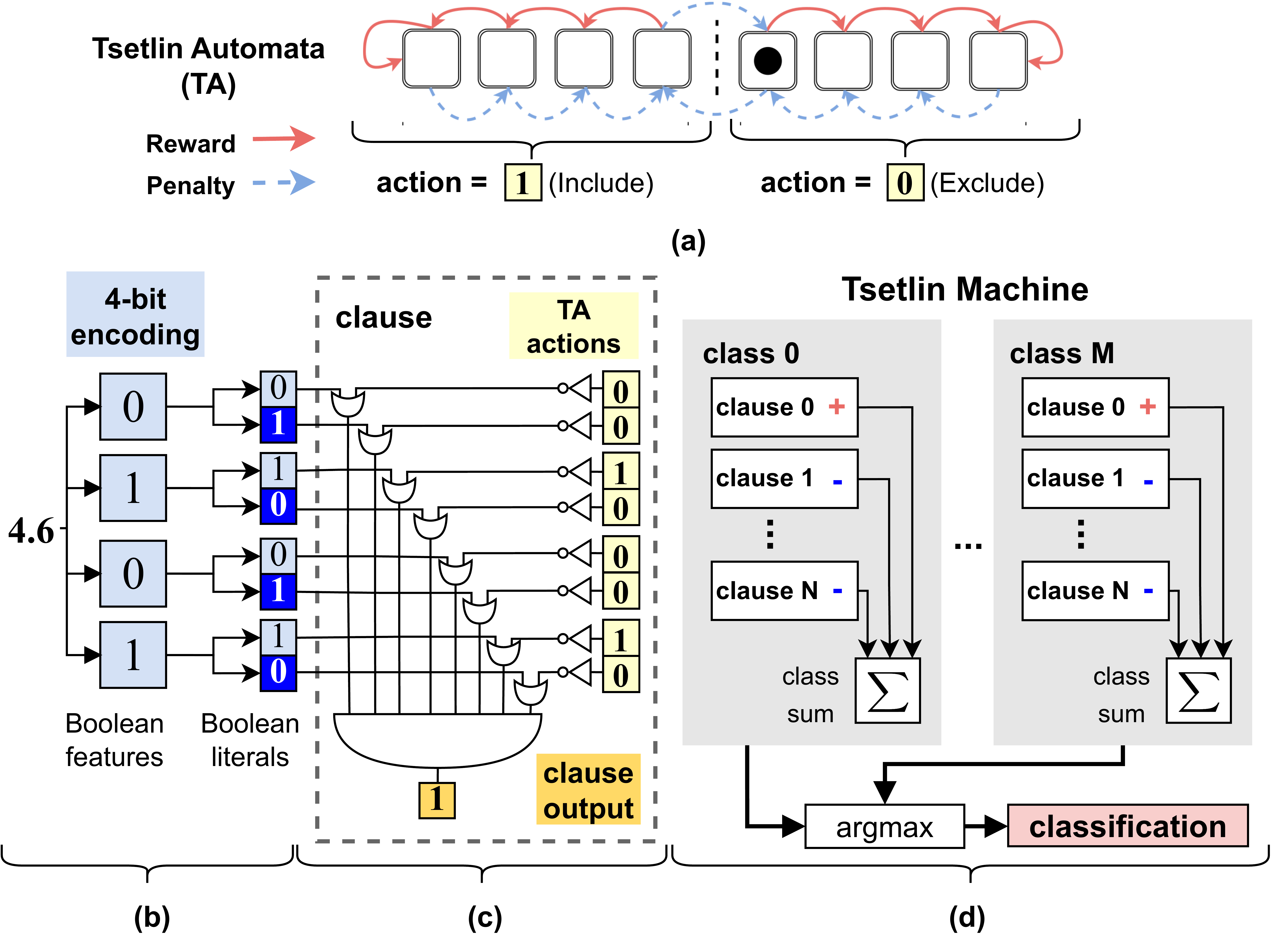}
    \caption{Diagram showing core components for inference using TMs: (a) the TA, (b) the booleanization of the input space, (c) the computation of a clause output and (d) the TM's architecture for inference.\vspace{-3mm}}
    \label{fig:TM_inf}
\end{figure}

While metal-oxide based ReRAM crossbars provide the means for both computation and storage through current summations~\cite{Xia2019,Yu2018,Ni2017}, when considering the wider ML inference procedure, there is a significant logic expense required to supplement the ReRAM architectures in the form of high bit precision ADCs and DACs~\cite{Wu2019} or through increased control logic complexity~\cite{Ni2017}.


This paper explores an alternative approach to tackling the above design challenges through an inherently bitwise inference algorithm called the \textbf{Tsetlin Machine (TM)}, see Fig.~\ref{fig:TM_inf}. The TM structure utilizes learning units called Tsetlin Automata (TA) to form logic propositions that relate Boolean inputs to the output for classification tasks without the need for multiply accumulate units. The simplicity of this algorithm is developed into an ReRAM crossbar to propose an In-Memory Boolean-to-Current Inference Architecture (IMBUE). A brief overview of the inference algorithm for the TM is given in the following paragraphs, and a more in-depth expansion of the algorithmic concepts and hardware-oriented nature of the TM can be found in~\cite{Granmo2018, wheeldon2020learning, Hotmobile22}.     



\textbf{Tsetlin Automata}: The TAs are the learning units that perform the classification and decision-making tasks in TM (Fig.~\ref{fig:TM_inf}a). Each automaton behaves like a finite state machine where half of the states of the TA are dedicated to an action. Reward and penalty stimuli move the automaton between the two actions: include or exclude. During training, rewards and penalties are issued to find optimal state positions for each TA. Once the training is completed, the TA can be viewed as one of the two possible Boolean values `1' and `0', indicating whether to include (`1') or exclude (`0') a particular feature. The input features must also be brought to a `1'/`0' Boolean domain to use this concept for feature selection. In doing so, it introduces two fundamental aspects of the TM: a) the booleanization of the input space, and b) how the logical propositions called clauses can be constructed to relate these Boolean inputs to the TAs.      


\textbf{Booleanization:} Fig.~\ref{fig:TM_inf}b shows how a raw value `4.6' can be converted to a 4-bit encoding to form Boolean features. Each Boolean feature is then further broken into Boolean literals which are the features and their complements. Now, these literals may interact with their respective TAs to compute a clause. While Fig.~\ref{fig:TM_inf}b has only one raw value, real TM applications will booleanize all the raw input values to compute a clause output.

\textbf{Clause Computation:} Fig.~\ref{fig:TM_inf}c shows how each Boolean literal can be related to the actions of each dedicated TA through NOT, OR and AND logic to form a 1-bit clause output. The TM contains many clauses, each with its own set of TAs that create different logic propositions from the same Boolean literals. The TM organizes these clauses into an architecture that enables classification.   

\textbf{TM Architecture:} For multi-class problems, the TM contains a fixed and equal number of clauses for each class. Within these classes, each clause can also have a polarity. Through the training process, the clauses with positive polarity learn the logic that supports the classification of their class. Clauses with negative polarity learn logic propositions that oppose their class. This is seen through the $+$ and $-$ signs in the clause blocks in  Fig.~\ref{fig:TM_inf}d. Each class has an equal number of positive and negative polarity clauses. Clauses with positive polarity have their outputs multiplied by (+1), while negative polarity clauses are multiplied by (-1). 

\textbf{TM Inference:} The TM's inference process is outlined by assembling all the above elements. First, the input Boolean literal is seen by all clauses in all the classes of the TM. Each clause will output a 0/1 based on its TA actions. Then the clause outputs will be multiplied according to their polarities and summed for each class. The $argmax$ of these class sums is the predicted classification.

Aside from the simplicity of the clause computation, one key aspect of the TM is the sparsity of TAs with include actions compared to exclude actions once the TM is trained. Upon viewing the clause computation, it is clear to see that only include TAs have an impact on the final clause output. This paper seeks to exploit this behavior through a representation of TA actions in ReRAM as either high and low resistances and the representation of input Boolean literals as Boolean voltages help to realize the Boolean-to-Current cells where the interaction between these literals (V) and these TAs (R) is further simplified into two distinct currents (I) ranges that may be summed according to the principles of Kirchhoff's Current Law (KCL). This forms the basis of the IMBUE Boolean-to-Current mechanism.

This paper proposes IMBUE for an ReRAM crossbar architecture to model the TM inference process. It investigates the non-linearity in current conversion, ReRAM device variations and evaluates the energy efficiency of the full system. In doing so, the following contributions are offered:
\begin{itemize}[]
\item for the first time, integration of 1T1R ReRAM technology into a new ML inference method utilizing Boolean-to-Current paths by following KCL;
\item investigation into variation tolerance of the TM under Device-to-Device (D2D), Cycle-to-Cycle (C2C) and CMOS variations; and
\item extensive evaluation of the efficiency of IMBUE in terms of power, latency and accuracy compared to it's closest Binary Neural Network and Digital TM state-of-the-art counterparts.
 
\end{itemize}

Next, Section II presents IMBUE, Section III presents experimental results and validation for the design, Section IV examines the energy efficiency of the proposed inference system and Section V concludes the paper.



\section{Proposed IMBUE Method} \label{proposed_design}
\begin{figure}[!t]
    \centering
    \includegraphics [width=\linewidth]{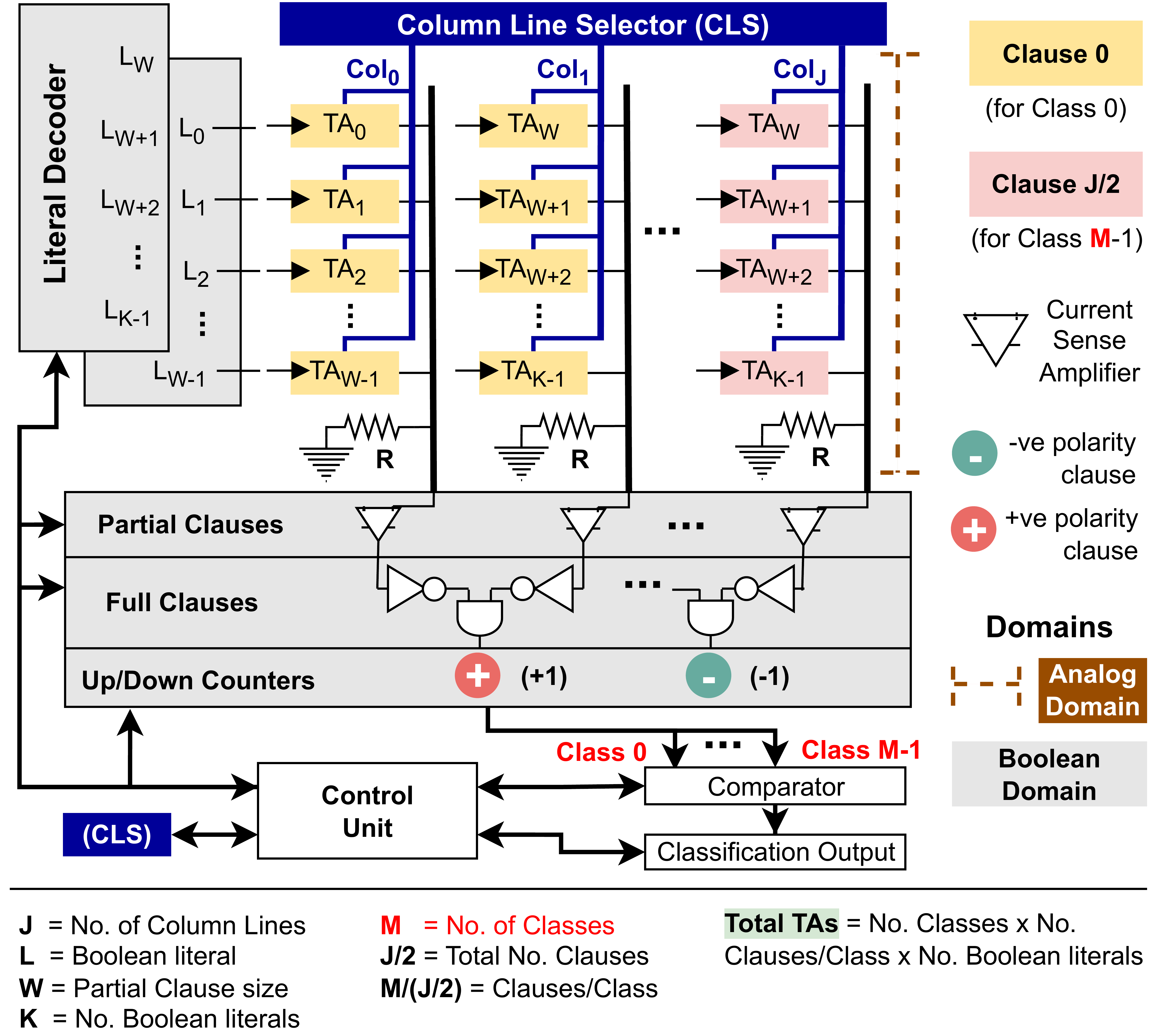}
    \caption{Proposed TM inference architecture using IMBUE.\vspace{-1mm}}
    \label{fig:ARRAY}
\end{figure}

The principal design concept of this paper is the ReRAM crossbar architecture for the TM inference presented in Fig.~\ref{fig:ARRAY}. At its core, the crossbar uses IMBUE for Boolean-to-Current conversion to realize clause computation. This section builds the TM inference procedure presented in Fig.~\ref{fig:TM_inf}, first focusing on mapping the TA as a ReRAM cell and creating clause outputs, then building to the full classification process.

\begin{figure}[!t]
    \centering
    \includegraphics[width=\linewidth]{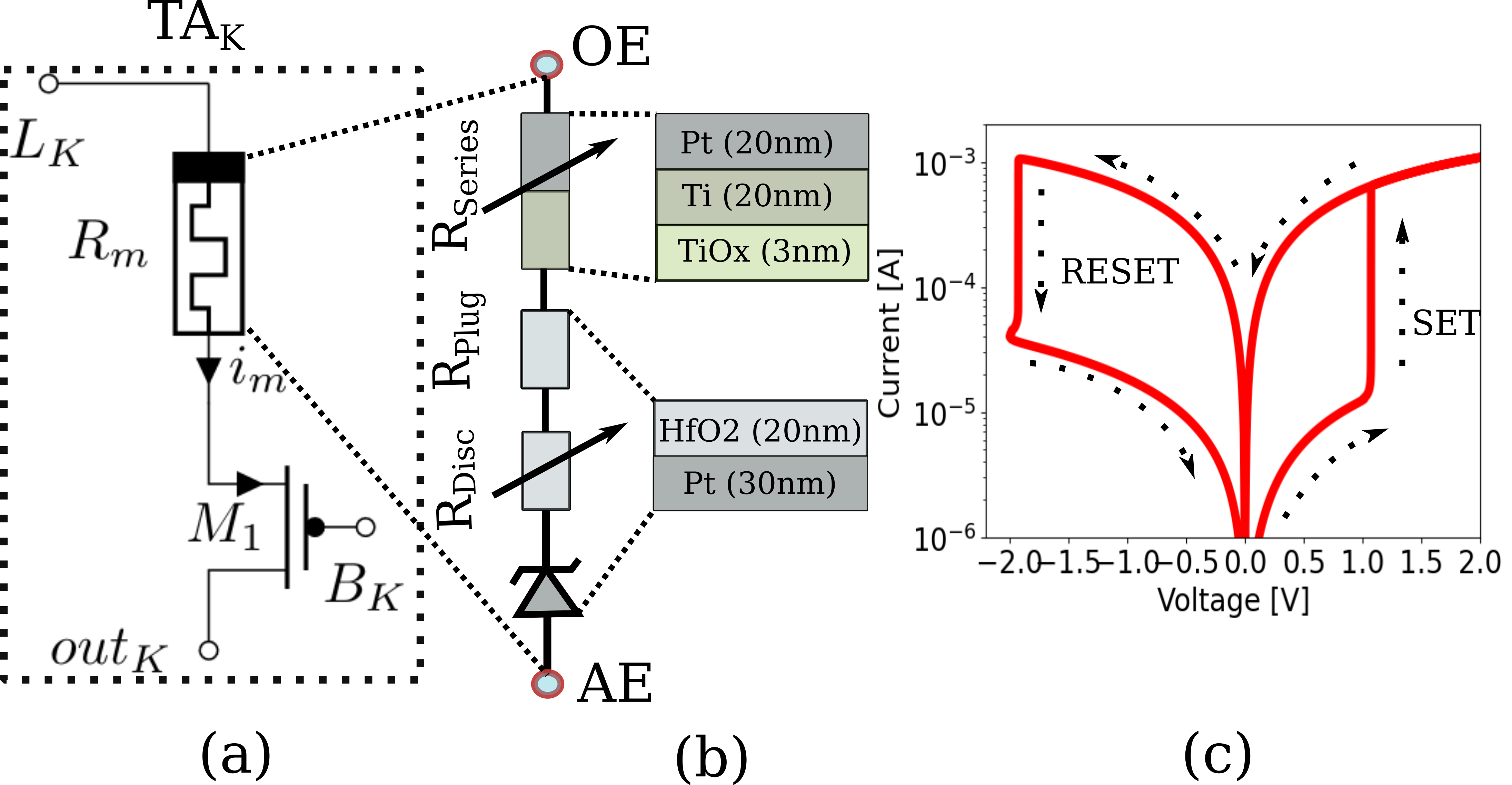}
\caption{Equivalent circuit for a TA in (a) represented by 1T1R cell, (b) shows the material stacks of memristor along with the I-V characteristics in (c). \vspace{-3mm}}
\label{1t1r_arch}
\end{figure}

\subsection{TA Cell using 1T1R}\label{1T1R}
The design of the TA cell is constructed with one memristor and one PMOS transistor forming the 1T1R structure. Fig.~\ref{1t1r_arch}a shows the architecture this design along with the equivalent stack of the memristor in Fig.~\ref{1t1r_arch}b. The memristor model used in this study consists of a Pt/Ti/TiO$_x$/HfO$_2$/Pt material stack, which is proven to have good electroforming voltage stability and  thermal stability~\cite{Benge , Hardtdegen}. Fig.~\ref{1t1r_arch}c shows the I-V characteristics of this memristor device, which has only two states, High Resistive State (HRS) = logic `0' used for TA exclude action and Low Resistive State (LRS) = logic `1' for the TA include action. The Boolean-to-Current mechanism is formed by the Ohm's Law relation when the literals, now seen as input voltages, are dropped across the TA resistances to produce a current. Fig.~\ref{fig:ARRAY} shows these TAs organized into columns. These columns are partial clauses which constitute only part of the full clause computations. This is done to minimize the effect of sneak currents and device variations. The partial clauses are examined next. 

\subsection{Partial Clause Computation}\label{Partial_clause_sec}
The partial clause evaluates part of the full clause using the Boolean-to-Current mechanism. 
In Fig.~\ref{fig:ARRAY} one full clause needs $K$ Boolean literals and respective TAs. This is divided into two partial clauses that now compute $K/2$ Boolean literals with their TAs (seen in yellow). The mapping of this split is done through the literals decoder such that all $K$ Boolean literals interact with their corresponding TAs for every clause. This is done for all clauses for all classes in the TM (the last partial clause of the last class is seen in pink). To perform a particular full clause computation the column line selector activates the two respective partial clause columns.    

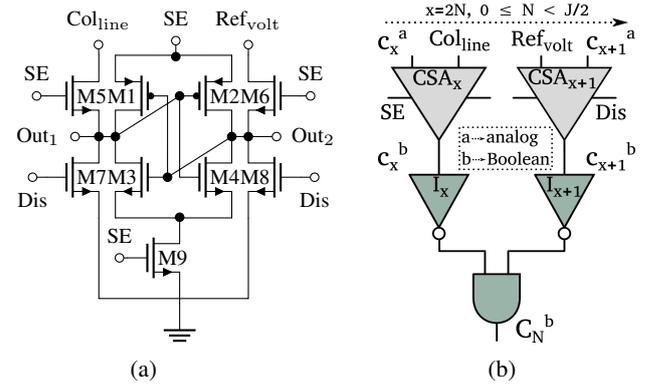
\begin{figure}[!t]
    \centering
    \begin{subfigure}{0.45\columnwidth}
    \centering
    \begin{circuitikz}[scale=0.53]
    \footnotesize
        \draw
        (-1.5,0) ++(3,0) node[nmos, anchor=drain,xscale=-0.7,yscale=0.7] (M3) {}
        (M3.drain) node[pmos, anchor=drain, xscale=-0.7,yscale=0.7] (M1) {}
        (M1.gate) ++(0.3,0)node[pmos, anchor=gate,xscale=0.7,yscale=0.7] (M2) {}
        (M2.drain) node[nmos, anchor=drain,xscale=0.7,yscale=0.7] (M4) {}
        (M1.gate) ++(-3,0)node[nmos, anchor=gate,xscale=0.7,yscale=0.7] (M5) {}
        (M3.gate) ++(-3,0)node[nmos, anchor=gate,xscale=0.7,yscale=0.7] (M7) {}
        
        (M2.gate) ++(3,0)node[nmos, anchor=gate,xscale=-0.7,yscale=0.7] (M6) {}
        (M4.gate) ++(3,0)node[nmos, anchor=gate,xscale=-0.7,yscale=0.7] (M8) {}
        (M8.source) ++(-3.0,-1.0)node[nmos, anchor=gate,xscale=0.7,yscale=0.7] (M9) {}

        (M3.gate) to [short] (M1.gate)
        (M2.gate) to [short] (M4.gate)
        (M3.source) to [short] (M4.source)
        (M1.drain) to [short,*-] (M5.source)
        (M1.source) to [short] (M2.source)
        (M3.source) ++(1.475,0) to [short, -*] (M9.drain)
        (M7.source) |- (M9.source)
        (M8.source) |- (M9.source)
        
        (M1.drain) [short, *-] (M2.gate)

        (M5.gate) + (-0.25,0) node[label=above:SE] {} to [short, o-] (M5.gate)
        (M6.gate) + (0.25,0) node[label=above:SE] {} to [short, o-] (M6.gate)
        (M7.gate) + (-0.35,0) node[label=below:Dis] {} to [short, o-] (M7.gate)
        (M8.gate) + (0.35,0) node[label=below:Dis] {} to [short, o-] (M8.gate)
        (M9.gate) + (-0.2,0) node[label=above:SE] {} to [short, o-] (M9.gate)
        
        (M1.source) (3.,2.4) node[label=above:SE] {} to [short, o-*] ++(0,-0.4)
        
        (M5.drain) (1.08,2.3) node[label=above:Col$\rm _{line}$] {} to [short, o-] (M5.drain)
        
        (M6.drain) (4.8,2.3) node[label=above:Ref$\rm _{volt}$] {} to [short, o-](M6.drain)
        
        (M8.drain) (5.5,0) node[label=right:Out$_2$] {} to [short, o-*] (M8.drain)
        
        (M5.source) (0.50,0) node[label=left:Out$_1$] {} to [short, o-*] (M5.source)
        
		(M2.drain) to [short, *-] (M6.source)
		(M2.gate) to [short, *-] (M1.drain)
		(M3.gate) to [short, *-] (M4.drain)
       
        (M9.source) node[ground] {}

        (M3.gate) node[left = 3mm]{M3}
        (M1.gate) node[left = 3mm]{M1}
        (M2.gate) node[right = 3mm]{M2}
        (M4.gate) node[right = 3mm]{M4}
        (M5.gate) node[right = 3mm]{M5}
        (M7.gate) node[right = 3mm]{M7}
        (M6.gate) node[left = 3mm]{M6}
        (M8.gate) node[left = 3mm]{M8}
        (M9.gate) node[right = 3mm]{M9}
        ;
    \end{circuitikz}
  \caption{\vspace{2mm}}
    \label{fig:CSA_circuit}
    \end{subfigure}\hfill 
\begin{subfigure}{0.5\columnwidth}
  \centering
  \includegraphics[width=0.75\linewidth]{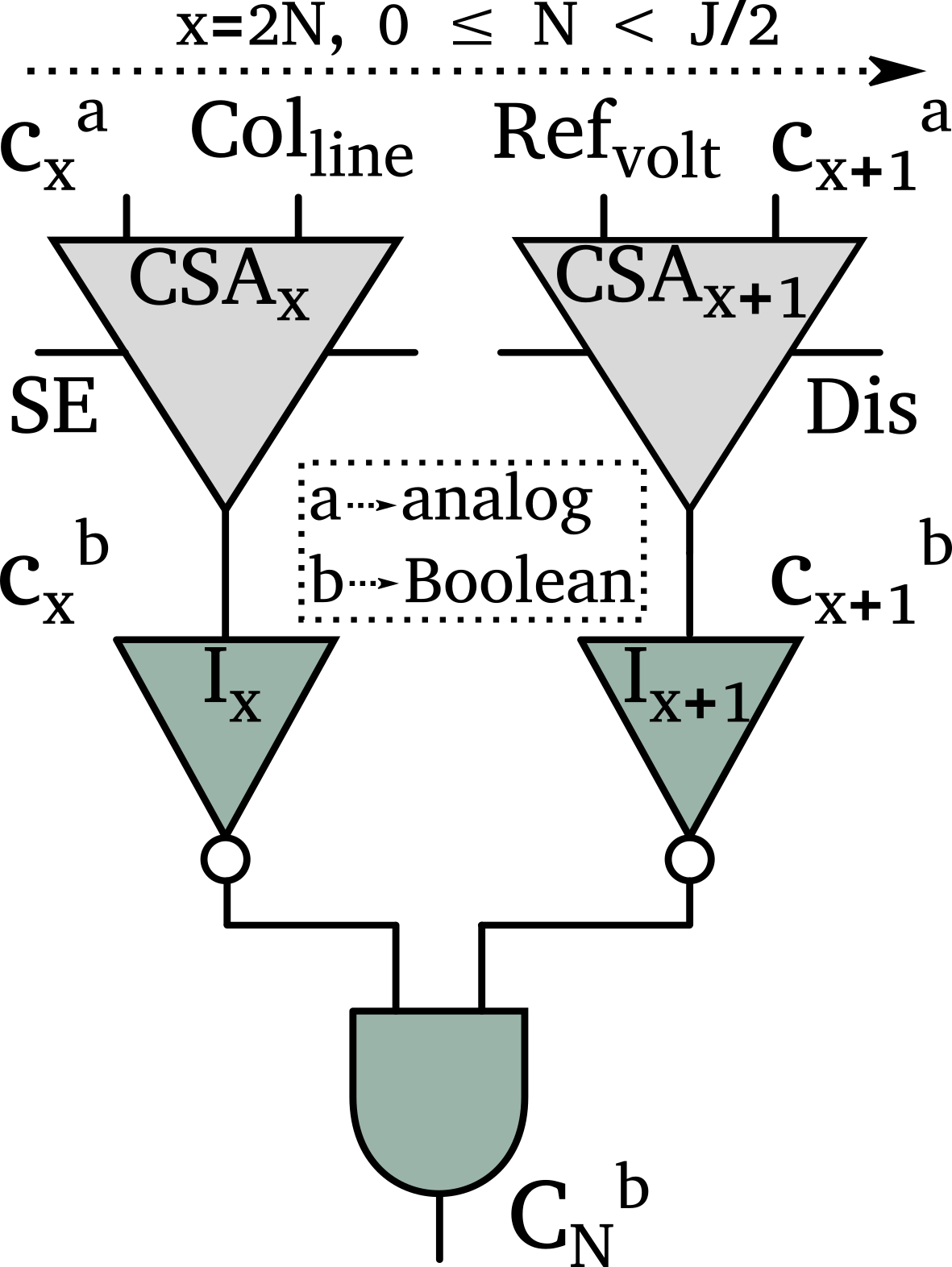}
  \caption{\vspace{2mm}}
  \label{fig:Clausemapping}
\end{subfigure}\hfill
    \caption{ CSA to convert current to Boolean in (a) and construction of a full clause using partial clauses is given in~(b).  \vspace{-3mm}}
\end{figure}

At the end of each column is an $R$ resistor. This is used to convert the KCL current to a column voltage Col$\rm _{line}$ in Fig.~\ref{fig:ARRAY}. Fig.~\ref{fig:CSA_circuit} shows the Current Sense Amplifier (CSA) used to compare this Col$\rm _{line}$ voltage with a Ref$\rm _{volt}$ reference voltage to generate a full-swing rail-to-rail Boolean output. Taking into account the variation of memristor devices, careful design choices must be made to have a larger current range between having at least one include TA in a column and only exclude TAs. Therefore, the proposed CSA utilizes a 65 nm CMOS technology that can function with input voltages close to the ground with minimal variations and low supply voltage (1.2V). From the CSA structure in Fig.~\ref{fig:CSA_circuit}, M1-M4 transistors combine to form two cross-coupled inverters of equal size and the input Col$\rm _{line}$ and Ref$\rm _{volt}$ are applied to M5 and M6, respectively.
During the partial clause evaluation (reading phase), SE is connected to VDD, allowing the CSA to determine which line has the higher voltage and produces a near Boolean'1' or '0' full swing output to the partial clause. Two additional 1.2 µm NMOS transistors M7 and M8 are used to discharge the two internal sensing nodes Out$_1$ and Out$_2$ to avoid adding a bias to subsequent CSA voltage comparisons. These partial clauses are then combined into full clause outputs.
\subsection{Full Clause Computation }\label{full_clause_sec}
Dividing into partial clauses helps minimize the non-linearity of the 1T1R cell's behavior. As the number of TAs within a clause increases, the variations in current contributions of HRS TA cells (exclude actions) may become large enough to equal an LRS TA cell (include) or affect the CSAs voltage comparison range. In Fig.~\ref{fig:Clausemapping} the inverters and AND gate compute a full clause from the partial clause outputs given by $C_{N}= c_{2N} \wedge c_{2N+1}$, \rm where $ N=0\rightarrow J/2$, $c_{2N}=\sum_{i=0}^{W-1}V(TA_{i}(I)),and  \ c_{2N+1}=\sum_{i=W}^{K-1}V(TA_{i}(I))$. \(V( TA_{i}(I))\) which indicates the Col$\rm _{line}$ voltage sensed by the CSA as a result of the KCL from that respective TAs column.

\subsection{TM System Level Inference}\label{TM_system_level}
As seen with Fig.~\ref{fig:TM_inf}(d)
full clauses must be assigned a polarity before contributing to a class sum. The architecture in Fig.~\ref{fig:ARRAY} achieves this through up/down counters. The $m$ class sums are then considered through a comparator to produce the final classification. These components are tied together through the control unit.


\section{Experimental Results and Validations}\label{Experimental}
This section evaluates the efficiency of IMBUE. We will study the characteristics of the TA cell under writing and reading operations. Later, we will also examine the modularity and scalability of IMBUE using the Boolean-to-Current mechanism and the CSA timing considerations. 
\subsection{1T1R TA Cell Characterization}\label{1T1R_chara_experiement}
\begin{figure}[!t]
    \centering
    \includegraphics [width=\linewidth]{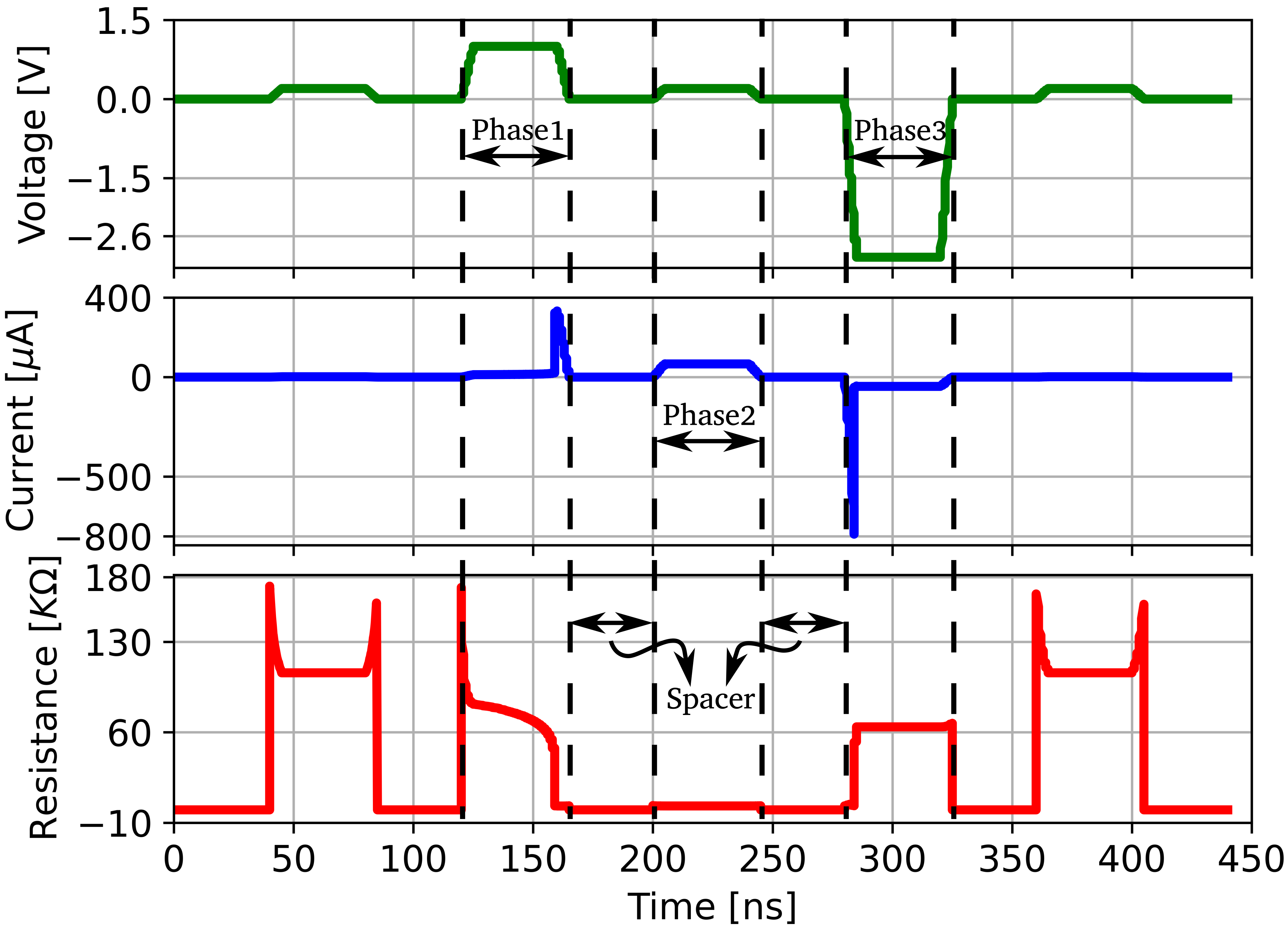}
    \caption{Programming and computation phases for a single TA.}
    \label{fig:VCR}
    \vspace{-3mm}
\end{figure}

 Initially, the TA cells must be programmed with the appropriate actions found through training. A voltage pulse with a specific duration and amplitude is applied to program the action in TA. Fig.~\ref{1t1r_arch}c shows the I-V characteristics of a device where SET (programming to include) and RESET (programming to exclude) paths are marked with arrows.  Table~\ref{tab:1T1R} shows the mapping of literal and action values on 1T1R cell in terms of reading voltage, resistance, and current respectively.
\paragraph{Programming action value in a TA cell}
TA programming involves two different programming voltage values named  $\rm V_{set}$ (1V) and $\rm V_{reset}$ (-2.5V). $\rm V_{set}$ changes the action from exclude to include, while $\rm V_{reset}$ changes it from the include to exclude value. A pulse duration of 35 ns is used to store these TA actions. The mechanism for programming any $TA$ cell can be done by activating the column line as the select column and disabling the CSA by setting SE to low. Finally, this voltage is applied to the required TA. 
$\rm V_{set}$ occurs during phase 1 in Fig.~\ref{fig:VCR} with a spark of rising current at the end of the applied pulse when the current is less than $0.4\ mA$. At the same time, the resistance value gradually decreases until, at the last few nanoseconds, a substantial drop coincides with the current spark reaching the maximum value. Phase 3 describes the reset operation, using $\rm V_{reset}$. During this phase, the current drops rapidly while the resistance reaches its expected value.
To provide a discharge period for the CSA, a Spacer of $0 V$ is used between the phases. The programming of TA actions is a one-time process. The programming combinations are described below:
\begin{itemize}
    \item \textit{Exclude} $\rightarrow$ \textit{Include}, applying $\rm V_{set}$ to switch to LRS.
    \item \textit{Include}  $\rightarrow$ \textit{Include}, applying $\rm V_{set}$ to stay in LRS.
  \item \textit{Exclude}  $\rightarrow$  \textit{Exclude}, applying $\rm V_{reset}$ to stay in  HRS.
  \item \textit{Include}  $\rightarrow$  \textit{Exclude}, applying $\rm V_{reset}$ to switch to HRS.
\end{itemize}

\begin{table}[!b]
    \centering
    \caption{Translate the 1T1R operation into TA concept.}
    \label{tab:1T1R}
    \setlength{\tabcolsep}{3pt}
    \begin{tabular}{|c|c|c|c|c|}
    \hline
         \makecell{Literal (logic) }& \makecell{Read (V)} & Action &  \makecell{Resistance ($K\Omega$)} & \makecell{Current ($\mu A$)}  \\
         \hline
         `0' & 0.2 & Include& $\approx$2.5 & $\approx$76.07   \\
         `0' & 0.2 & Exclude& $\approx$105.8 & $\approx$1.89\\
          `1' & 0 & Include& $\approx$7.6 & $\approx$137$e^{-9}$  \\
          `1' & 0 & Exclude& $\approx$33.6 & $\approx$9.9$e^{-9}$ \\
         \hline
    \end{tabular}
    
\end{table}

\begin{figure}[!t]
    \centering
    \includegraphics[width=\linewidth]{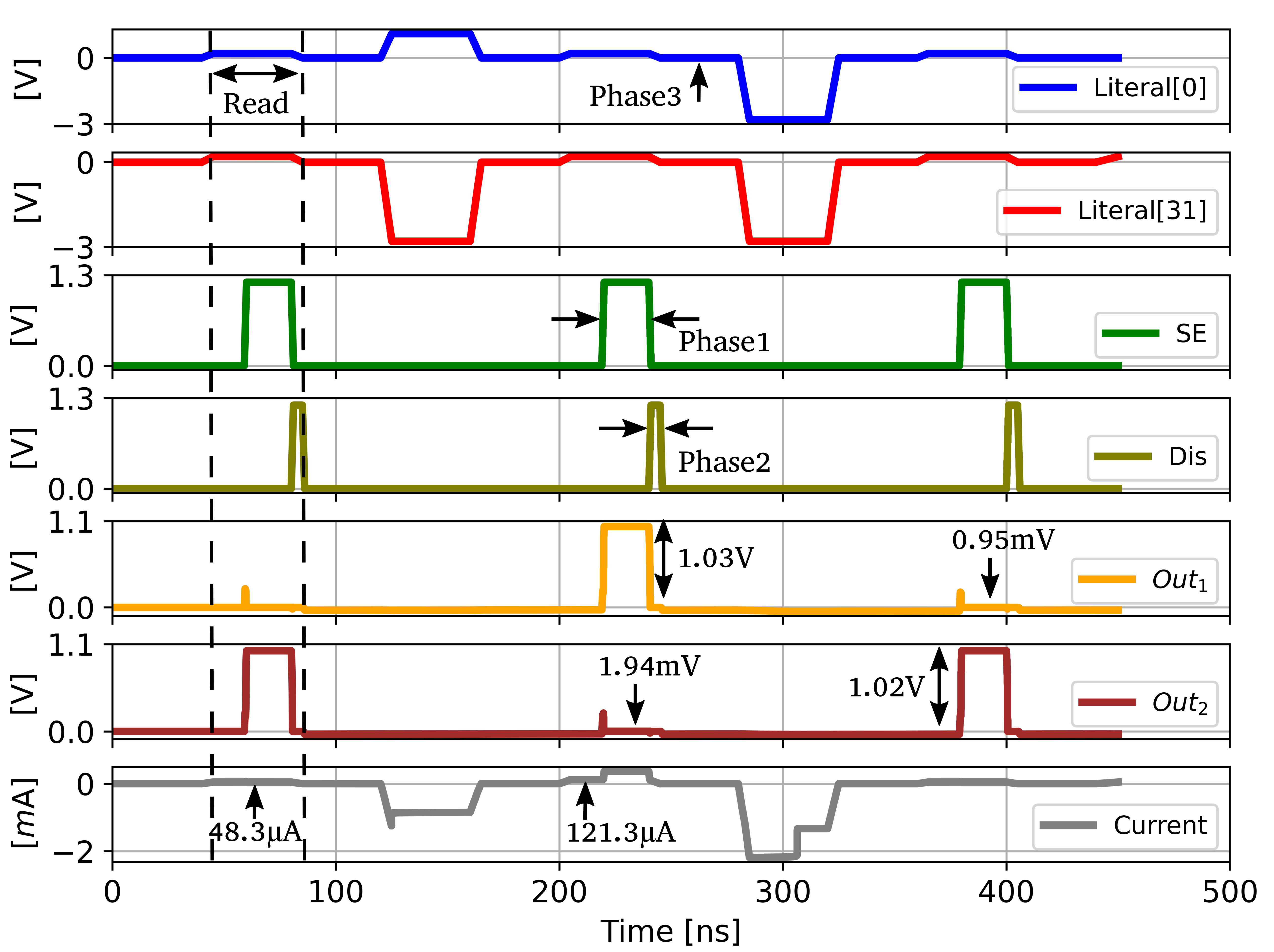}
    \caption{CSA waveforms in 3 phases. \vspace{-3mm}}
            \label{fig:CSAwaveform}
\end{figure}

\paragraph{Boolean-to-Current computation}
The Boolean literal input voltages are given as logic `1' $\rightarrow$ $0\ V$ and logic `0' $\rightarrow$ $0.2\ V$ as seen in Fig.~\ref{fig:VCR}. They are then applied to the selected partial clause column. The target TAs' Col$\rm _{line}$ is turned on so that the potential difference across the TA is somewhere between the two programming voltages.According to Ohm's law, the Boolean literal value and TA action interact to produce a proportional current. At the end of each Col$\rm _{line}$, a voltage divider resistance $(R=100\ \Omega)$ converts the output current to a voltage. This voltage is then sensed using the CSA as shown in the waveform Fig.~\ref{fig:CSAwaveform}. Table~\ref{tab:power_consumption} shows power consumption for all possible TA combinations. The goal is to dominate the \textit{include} with \textit{literal `0'} combination while maintaining the accuracy for less power consumption. This can be done in TM training by reducing the include/exclude ratio as much as possible. 







\begin{table}[!b]
    \centering
    \caption{Power consumption for 1T1R cell.}
    \label{tab:power_consumption}
    \begin{tabular}{|c|c|}
    \hline
         Operation & \makecell{TA power consumption ($\mu W$)} \\
         \hline
         Program to \textit{Exclude} & 54.54 \\
         Program to \textit{Include} & 215.1 \\
         \textit{Include} $\times$ Literal `0' & 14.37 \\
         \textit{Exclude} $\times$ Literal `0' & 377.2$e^{-3}$ \\
         Otherwise & $\approx 0$ \\ 
         \hline
    \end{tabular}
\end{table}

\subsection{Partial Clausing through CSA Operation} 
\label{partial_clausing}

The current created by the partial clause is outlined in three operating phases. The first phase occurs during the Col$\rm _{line}$ reading pulse (35 ns), and SE is held to VDD (20 ns) to enable the latch. The voltage across the column resistor  R is sampled and supplied into the CSA's internal nodes Out$_{1}$ and Out$_2$. The higher voltage node is set to a higher output voltage near VDD, whereas the lower voltage node is set near GND. In the second phase, SE is set to GND, while Dis is set to VDD for a (5 ns) spark to discharge the two output nodes to the ground voltage level. The SE and Dis signals are grounded in the final phase to prepare the CSA for the next sensing cycle. The waveforms for the CSA design at 1.2V VDD connected to a $\rm Col_{line}$ of 32 TAs is shown in~Fig.~\ref{fig:CSAwaveform}.



\subsection{Impact of Variations} \label{variation}
In this section, we study how manufacturing process variations affect the design choices of  architecture. First, we discuss the variation in ReRAM cells followed by variation in CMOS-based components in the architecture, especially CSA.
\subsubsection{ReRAM cell variations}
The D2D and C2C variations can affect the switching behavior of ReRAM devices. The following are the parameters affected during the manufacturing process of ReRAM used in this study. The parameters' ranges used in this study are the same as in~\cite{Benge}:
\begin{itemize}
    \item Maximum and minimum oxygen vacancy concentration in the disc, ($N_{Disc,max}$ and $N_{Disc,min}$).
    \item Radius and length of the disc, ($r_{disc}$ and $l_{disc}$).
\end{itemize}

\paragraph{C2C variation} 
C2C is analyzed by altering the variable parameters before each SET and RESET cycle. A MATLAB script randomly increases or decreases (with equal probability) each parameter's value from its previous value every cycle. These are then exported as a CSV file. This file contains the varying parameters and their timestamps and is used for the Spectre simulation in this study.
\begin{figure}[!t]
    \centering
    \includegraphics[width=\linewidth]{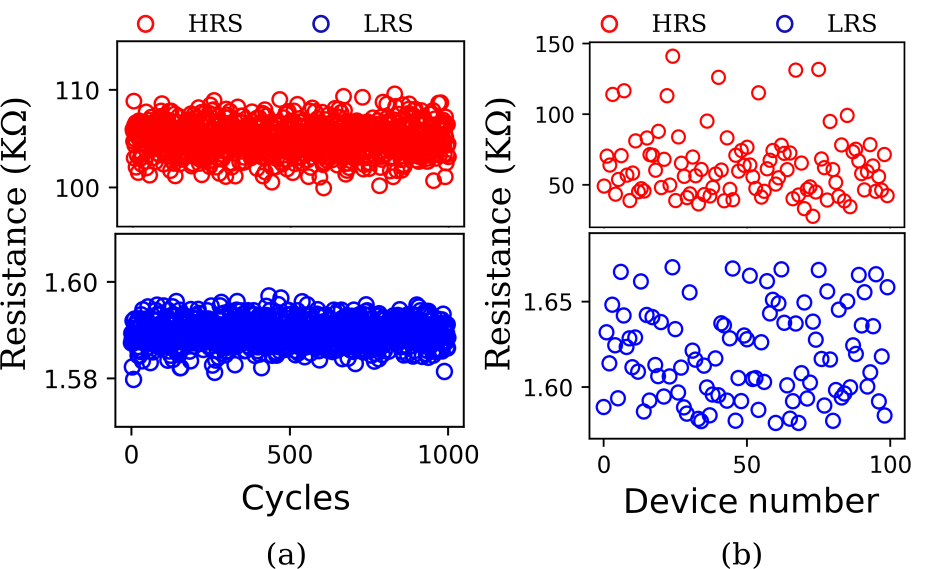}
\caption{LRS and HRS distribution for 1000 cycles under C2C variations in (a) D2D variation on 10x10 crossbar in (b). \vspace{-3mm}}
    \label{fig:variations}
\end{figure}
Fig.~\ref{fig:variations}a shows the distribution of HRS and LRS for 1000 cycles. The new set of values is applied after each 100 ns period. Within these 100 ns, the device is RESET with 25 ns (20 ns pulse width with 2.5 ns rise and fall time) pulse of -2 V and SET with 2 V pulse with the same duration. The state of the device is read by applying a READ pulse of 0.2 V after every RESET/SET operation. In the next cycle, new values are 
applied in the simulation. It has been observed that there is a $\pm$5\% change in HRS and an $\approx$1\% change in LRS during C2C variations.

\paragraph{D2D variation} 
The experimentally validated Gaussian distribution from ~\cite{Benge} is used to generate a random set of values for each device's varying parameters. These variations were then individually applied to the experiment's devices. The distribution of HRS and LRS states in a 10x10 crossbar without a transistor is shown in Fig.~\ref{fig:variations}b. To measure the HRS distributions, all devices are first switched to LRS by applying a 2 V SET pulse with a duration of 150 ns and a rise and fall period of 50 ns. The devices are switched to the HRS state in the next cycle by applying a -2 V RESET pulse. At 0.2 V, the device's resistance status is read. The same method is used to measure LRS distribution. HRS has been found to have a greater impact on variance than LRS. Due to variation, the HRS vary from 31-155\ $K\Omega$ with an average of 65.56\ $K\Omega$ and LRS from 1.55-1.67\ $K\Omega$ with an average of 1.64\ $K\Omega$. The ratio of HRS and LRS decides how many TAs can be fitted into a single column to compute partial clauses accurately.

\begin{figure}
    \centering 
    \includegraphics[width=\linewidth]{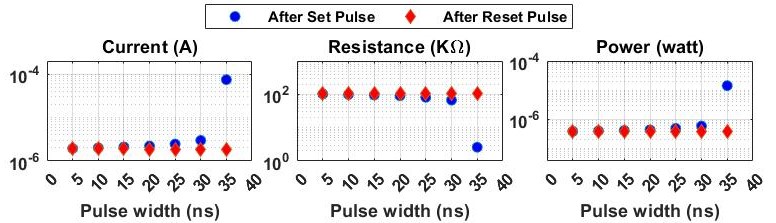}
    \caption{Current, Resistance and Power variation for different programming pulse duration.\vspace{-3mm}}
    \label{fig:Diff_Periods}
\end{figure}


\begin{table}[!b]
\setlength{\tabcolsep}{2pt}
    \centering
    \caption{CSA's corners and process variation analysis.}
    \label{tab:CSAcorners}
    \begin{tabular}{|c|c|c|c|c|c|c|c|c|}
    \hline
   & & \multicolumn{5}{c|}{Corner analysis} & \multicolumn{2}{c|}{Process variations} \\
   \cline{3-9}
  Operation & CSA	&Nominal & FF	&SS	&SF	&FS	&Mean	&SD \\
  \cline{3-9}
    & output & \multicolumn{7}{c|}{(mV)} \\
    \hline
SET & Out$_1$ &865.7&860.3&867.1&892 &819.6 & 864.86 & 10.35 \\
SET & Out$_2$	&0.95	&3.95	&0.32	&1.32	&0.99 & 1.06 & 0.445 \\
RESET &Out$_1$ &28.35	&30.06	&28.33	&28.75	&28.38 &	28.44 & 0.216 \\
RESET&	Out$_2$ &876.4 &872.9	&874.7	&908.1	&823.1 &	875.46 & 12.33 \\
         \hline
    \end{tabular}
\end{table}

\begin{table*}[!t]
\centering
  \caption{Energy estimation and comparison to state-of-the-art TM for different datasets.}
    \label{tab:energy}
\begin{tabular}{|c|c|c|c|c|c|c|c|c|c|c|}
\hline
\textbf{Dataset} & \textbf{Accuracy } & \textbf{Classes} & \textbf{Clauses} & \textbf{TA cells} & \textbf{Includes} & \textbf{Includes} & \textbf{CSAs} & \textbf{CMOS TM}~\cite{wheeldon2020learning} & \textbf{IMBUE} & \textbf{x energy}  \\
\cline{9-10}
\textbf{name}& \%& \textbf{\#} & \textbf{total} & \textbf{\#} & \textbf{\#} &\textbf{\%}&\textbf{\#} & \multicolumn{2}{c|} {\textbf{Average energy/datapoint (nJ)}} & \textbf{ reduction} \\

\hline 

Noisy XOR                  & 99.2  & 2 & 12   & 576& 48 & 8.3 & 18& 0.0092 &\textbf{ 0.02}    & \textbf{0.36}  \\
MNIST                    & 96.48 & 10 & 2000 & 3136000& 18927 &0.6 & 98000& 50.01 & \textbf{13.9} & \textbf{3.597}  \\
KWS-6 & 87.1  & 6  & 1800 & 1357200 & 7990 &0.58 & 42413  & 21.64 &\textbf{5.91} & \textbf{3.66}  \\
K-MNIST                     & 88.6  & 10 & 5000 & 7840000& 31217 & 0.39& 245000& 125.03 & \textbf{26.47}   & \textbf{4.722} \\
F-MNIST                     & 87.67 & 10 & 5000 & 7840000& 25742 & 0.32& 245000 & 125.03 & \textbf{23.66}   & \textbf{5.283} \\
\hline

\end{tabular}
\vspace{-4mm}
\end{table*}



\subsubsection{Pulse duration study} 
This study gives the minimum pulse duration required to switch the 1T1R device from LRS to HRS and vice-versa. 
Fig.~\ref{fig:Diff_Periods} shows the current, resistance, and power of a single device on different pulse duration (5 to 100) ns while the voltage of the pulse is constant ($\rm V_{set}=1\ V$ and $ \rm V_{reset}=-2.5\ V$). It is shown in Fig.~\ref{fig:Diff_Periods} that the device switched from HRS to LRS at 35 ns pulse width. A pulse width greater than 35 ns will lead to more power consumption and latency of the design. 

\subsubsection{CMOS corner and process variation}

Corner analysis and Monte Carlo analysis were performed to evaluate the dependability of the CSA. The worst case scenario of only one include TA in a partial clause column of 32 TAs is used. Each cycle a new set of literals is applied to this column. Out$_{1}$ and Out$_2$ readings are taken from the CSA for each cycle (see Fig.~\ref{fig:CSA_circuit}). Table~\ref{tab:CSAcorners} lists the two output's voltages for the corners analysis and the process variation for 2000 cycles with a small standard distribution.

\section{Comparative Energy Efficiency}\label{Power}

The energy efficiency of the proposed IMBUE based architecture is simulated with a Python script using the power consumption values seen in Table~\ref{tab:power_consumption} and the timing presented in Fig.~\ref{fig:CSAwaveform}. The crossbar size was kept as 32 TAs per partial clause column. TMs were trained for datasets including Noisy XOR, MNIST, Kuzushiji-MNIST (K-MNIST), Fashion-MNIST (F-MNIST), and Google Keyword Spotting dataset with 6 keywords (KWS-6), which were booleanized using the method presented in~\cite{Lei2021}. The TAs from these models were extracted and used to simulate the proposed architecture and compare against a CMOS TM implementation presented in~\cite{wheeldon2020learning}. The results showed that the number of include TAs in each of these TM models significantly impacted the resulting energy efficiency when applied to the proposed IMBUE based architecture. As the complexity of the learning problem increases, the TM is unable to select as many key include actions for the TAs. However, this benefits IMBUE's Boolean-to-Current mechanism greatly as it reduces the contribution of the most power-intensive part of the proposed inference system. Table~\ref{Power} presents the energy/datapoint values obtained for the different datasets using IMBUE and CMOS TM. The results showed that IMBUE outperformed CMOS TM in terms of energy efficiency for all datasets except for Noisy XOR which has the highest include/exclude ratio. 



\begin{figure}
    \centering 
    \includegraphics[width=\linewidth]{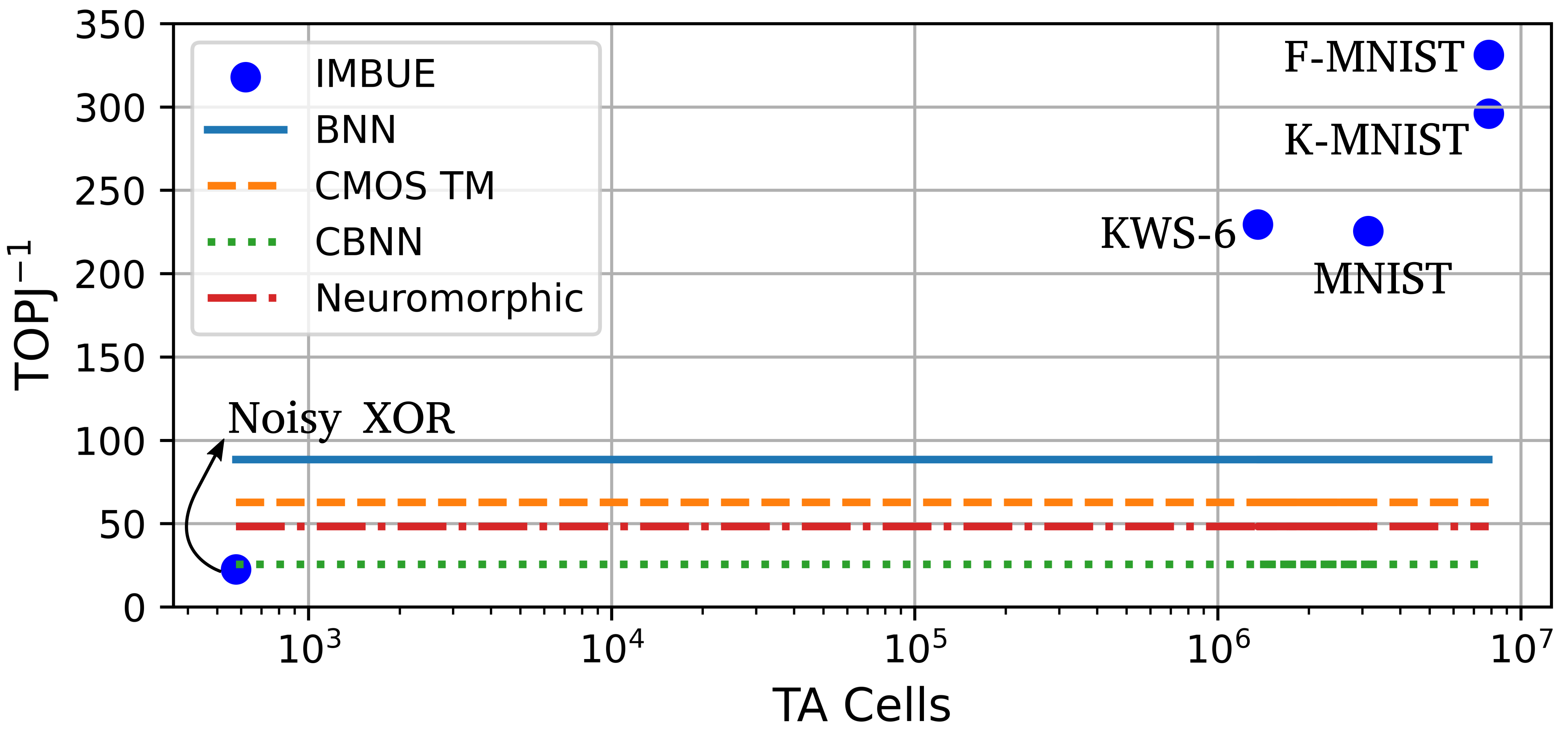}
    \caption{IMBUE energy efficiency for TM inference applied to various datasets.\vspace{-3mm}}
    \label{fig:TOPS}
\end{figure}

The closest comparable ML to TMs are Binary Neural Networks (BNNs) with both reliant on bitwise operations. However to avoid biases in the presence of algorithmic differences, we employed the TopJ$^{-1}$ metric, which measures trillion operations per second per joule of energy consumed. For TMs this can be calculated as the number of TAs used to compute all the clauses divided by the energy per datapoint. Fig.~\ref{fig:TOPS} shows how the proposed IMBUE architecture based TM models perform when compared against Neuromorphic~\cite{neuromorphic}, BNN~\cite{BNN} and Binarized Convolutional Neural Networks (CBNN)~\cite{CBNN}. As seen with Table~\ref{Power}, as the include sparsity increases for more complex TM models, the IMBUE solution offers better performance. Specifically, the TM inference based on IMBUE outperforms CMOS-based TM, BNN, CBNN, and Neuromorphic by up to 5.28x, 3.74x, 12.99x, and 6.87x, respectively, achieving a TopJ$^{-1}$ of 331 for F-MNIST. 

\vspace{-1mm}
\section{Conclusion}\label{Conclusions}
The IMBUE TM architecture was realized through a ReRAM crossbar and evaluated in terms of performance of the inference procedure and resilience against ReRAM variations. The core component of the TM, the Tsetlin Automaton, was implemented with a 1T1R cell. Each cell is programmed to either LRS or HRS to reflect the include or exclude TA actions respectively. The ReRAM resistances may vary upon reading, however D2D and C2C studies showed sufficient robustness in meeting the CSA margins when reading the stored TA actions. The rationale for IMBUE's Boolean-to-Current mechanism is made evident from the nature of the TM's clause computation; the logic based interaction between Boolean inputs and the TA include/exclude actions in the clause is simplified into current accumulation. To address the issues of sneak currents and non-linearity when converting these currents from the analog domain, the TM clause was broken into partial clauses. These partial clauses help mitigate device variations while still permitting scalability. The power consumption of the TM inference is dominated by two conditions as seen in Table~\ref{tab:power_consumption}. Examination of case study datasets exposes the rarity of such conditions. This is further seen through Table~\ref{Power} where increasing model complexity leads to increased sparsity in TM include decisions and therefore offers much better energy efficiency. For the F-MNIST dataset the proposed architecture gives 12.99x better TopJ$^{-1}$ compared with CBNN. 

The recent works with TMs have proposed coalesced clause architectures where clauses are shared between classes~\cite{CoalescedMT}. Future work aims to explore the associated trade-offs from applying the principles of IMBUE to such an algorithm.   

\section*{Acknowledgments}
The authors would like to gratefully acknowledge the funding support from the UK Northern Accelerator (ref: NACCF 220), Lloyds Registers Foundation (ref: 5thICON-12) and Norwegian Research Council (ref: AIEverywhere project).

\bibliographystyle{IEEEtran}
\bibliography{Bib}

\end{document}